\documentclass[usenatbib]{mnras}

\usepackage{xspace,amsmath}
\usepackage{color,epsfig,amssymb,lscape}
\usepackage{array,colortbl}
\usepackage{url}

\usepackage[T1]{fontenc}
\usepackage{ae,aecompl}

\newcommand{\hii}{\relax \ifmmode {\mbox H\,{\scshape ii}}\else H\,{\scshape ii}\fi}
\newcommand{\mi}{\relax \ifmmode {\mu{\mbox m}}\else $\mu$m\fi}
\newcommand{\ha}{\relax \ifmmode {\mbox H}\alpha\else H$\alpha$\fi}
\newcommand{\hb}{\relax \ifmmode {\mbox H}\beta\else H$\beta$\fi}

\newcommand{\sii}{\relax \ifmmode {\mbox S\,{\scshape ii}}\else S\,{\scshape ii}\fi}
\newcommand{\siii}{\relax \ifmmode {\mbox S\,{\scshape iii}}\else S\,{\scshape iii}\fi}
\newcommand{\nii}{\relax \ifmmode {\mbox N\,{\scshape ii}}\else N\,{\scshape ii}\fi}
\newcommand{\oi}{\relax \ifmmode {\mbox O\,{\scshape i}}\else O\,{\scshape i}\fi}
\newcommand{\oii}{\relax \ifmmode {\mbox O\,{\scshape ii}}\else O\,{\scshape ii}\fi}
\newcommand{\oiii}{\relax \ifmmode {\mbox O\,{\scshape iii}}\else O\,{\scshape iii}\fi}
\newcommand{\neiii}{\relax \ifmmode {\mbox Ne\,{\scshape iii}}\else Ne\,{\scshape iii}\fi}

\newcommand{\rdostres}{\relax \ifmmode {\,\mbox{R}}_{\rm 23}\else \,\mbox{R}$_{\rm 23}$\fi} 

\newcommand{\ciii}{\relax \ifmmode {\mbox O\,{\scshape iii}}\else C\,{\scshape iii}\fi}
\newcommand{\civ}{\relax \ifmmode {\mbox O\,{\scshape iii}}\else C\,{\scshape iv}\fi}
\newcommand{\nv}{\relax \ifmmode {\mbox N\,{\scshape v}}\else N\,{\scshape v}\fi}
\newcommand{\heii}{\relax \ifmmode {\mbox He\,{\scshape ii}}\else He\,{\scshape ii}\fi}

\newcommand{\gsim}{\hbox{\rlap{\lower.55ex\hbox{$\sim$}} \kern-.3em
\raise.4ex \hbox{$>$}}}
\newcommand{\lsim}{\hbox{\rlap{\lower.55ex\hbox{$\sim$}} \kern-.3em
\raise.4ex \hbox{$<$}}}

\usepackage{graphicx}	
\usepackage{amsmath}	
\usepackage{amssymb}	

\title[Model-based chemical abundances in AGN from UV]{Assessing model-based carbon and oxygen abundance derivation from ultraviolet emission lines in AGNs}

\author[E. P{\'e}rez-Montero et al.]{
E. P{\'e}rez-Montero$^{1}$\thanks{E-mail: epm@iaa.es (EPM)},
R. Amor\'\i n$^{2,3}$, B. P\'erez-D\'\i az$^1$,
J.M. V\'\i lchez$^1$, and
R. Garc\'\i a-Benito$^1$ \\
\\
$^{1}$Instituto de Astrof\'\i sica de Andaluc\'\i a. CSIC. Apartado de correos 3004. 18080, Granada, Spain.\\
$^{2}$Instituto de Investigaci\'on Multidisciplinar en Ciencia y Tecnolog\'ia, Universidad de La Serena, Raul Bitr\'an 1305, La Serena, Chile. \\
$^{3}$Departamento de Astronom\'ia, Universidad de La Serena, Av. Juan Cisternas 1200 Norte, La Serena, Chile. \\}

\date{Accepted XXX. Received YYY; in original form ZZZ}

\pubyear{2020}

\begin{document}
\label{firstpage}
\pagerange{\pageref{firstpage}--\pageref{lastpage}}
\maketitle

\begin{abstract}
We present an adapted version of the code {\sc HII-CHI-mistry-UV} \citep{pma17} to derive chemical abundances from emission lines in the ultraviolet, for use in narrow line regions (NLR) of Active Galactic Nuclei (AGN). We evaluate different ultraviolet emission line ratios and how different assumptions about the models, including the presence of dust grains, the shape of the incident spectral energy distribution, or the thickness of the gas envelope around the central source, may affect the final estimates as a function of the set of emission lines used. We compare our results with other published recipes for deriving abundances using the same emission lines and show that deriving the carbon-to-oxygen abundance ratio using \ciii] $\lambda$ 1909 \AA\ and \oiii] $\lambda$  1665 \AA\ emission lines is a robust indicator of the metal content in AGN that is nearly independent of the model assumptions, similar to the case of star-forming regions. Moreover, we show that a prior determination of C/O allows for a much more precise determination of the total oxygen abundance using carbon UV lines, as opposed to assuming an arbitrary relationship between O/H and C/O, which can lead to non-negligible discrepancies.
 \end{abstract}




\section{Introduction}

Active Galactic Nuclei (AGN) host one of the most powerful energy sources in the universe. 
The intense radiation field emanating from the hot accretion disks around supermassive black holes in galactic centers produces very bright emission lines from the surrounding gas. From this, various physical properties and chemical abundances in these regions can later be inferred up to a very high redshift, which can serve as indicators of the evolution of the universe at different cosmic epochs.

The UV region is of outstanding importance in this context. Several bright emission lines produced by AGNs, such as \nv, \heii, \civ, \oiii], and \ciii], which are sensitive to the ionisation conditions and physical properties (i.e., electron density and temperature) of the warm ISM \citep[e.g.,][]{Kewley2019}, can be identified in the $\lambda\sim$1000\AA\ -2000\AA\ region. While this range at $z\lesssim2$ is typically explored using the {\it Hubble Space Telescope} \citep[HST; e.g.][]{Rigby2018, Berg2022}, these lines are redshifted to rest-optical wavelengths at $z\sim$2-4. Therefore, deep optical spectroscopic surveys with ground-based telescopes of 8-10m class \citep[e.g.][]{Steidel2003,Shapley2003,zcosmos,Kurk2013,VUDS,VANDELS} are typically efficient at detecting the largest samples of UV line emitters \citep[e.g.][]{Steidel2014,Maseda2017,amorin17,Nakajima2018a,lefevre2019, Schmidt2021,Feltre2020, Saxena2020,Saxena2022,Llerena2022}. 

Classical methods for determining chemical abundance are instead mostly based on rest- optical emission lines \citep[e.g.][]{Maiolino2019}. However, for galaxies at $z\sim$\,2-3 these are redshifted to the near-infrared (NIR) and detections are often limited to the few bright lines \citep[ e.g.][]{Steidel2014,Shapley2015}. Joint analysis of the observed rest- UV and rest- optical emission spectra with predictions from detailed photoionisation models for both star-forming galaxies \citep[ e.g.][]{Gutkin2016,Byler2018,pma17} and AGNs \citep[ e.g.][]{Feltre2016,Nakajima2018b,dors19, Hirschmann2019,mignoli19} thus emerge as powerful diagnostics for the ionisation and chemical abundances of galaxies \citep[e.g.][]{ Patricio2016,Vanzella2016b,amorin17,Byler2020} and pave the way for similar analyses at higher redshifts with the {\it James Webb Space Telescope} \citep[e.g.][]{Chevallard2019,Rigby2021}. 

The derivation of the metal content in the Narrow Line Region (NLR) of AGN galaxies from optical collisionally excited lines (CELs)
is much more complex than in the case of star-forming regions, since the direct method (i.e., based on the prior determination of the electron temperature) can significantly underestimate the derived oxygen abundances \citep[e.g.][]{dors15},
which are often considered as proxies for gas metallicity in galaxies.
Instead, there are several studies dealing with the determination of different elemental abundances in AGN based on pure photoionisation models \citep[e.g.][]{storchi98,castro17,thomas19} or on models assuming shocks in combination with photoionisation \citep[e.g.][]{dors21}.

In the case of UV emission lines, since direct determination of chemical abundances in AGN is still not reliable, many of the various approaches pursued are also based on photoionisation models, including detailed models for individual objects \citep[e.g.][]{davidson77, osmer80, uomoto84, gaskell81, hamann92, ferland96, dietrich00, hamann02, shin13, feltre16, yang17}.
For a large number of objects, other authors propose calibrations of some UV emission line ratios sensitive to chemical abundance in these detailed models \citep[e.g.][]{dors14,dors19}, 
or compare them in model-based diagnostic diagrams \citep[e.g.][]{nagao06, matsuoka09, matsuoka18}.

Another strategy suitable for use in large surveys is to use a Bayesian-like comparison between the adequate emission line ratios and the predictions from large model grids \citep[e.g.][]{mignoli19}.
This method has the advantage of better quantifying uncertainties and easily specifying the assumptions required when the number of emission lines input is limited. 
For example, the use of photoionisation models, which include various combinations of
C/O and O/H has great significance for estimating gas-phase metallicity in cases where very few C lines (e.g., \civ $\lambda$ 1549 \AA \ and \ciii] $\lambda$ 1909 \AA) UV emission lines can be measured.

In this work, we present an adapted version of the code 
{\sc Hii-Chi-mistry-UV} (hereafter {\sc HCm-UV}, \citealt{pma17}),
originally developed for deriving O/H and C/O from UV emission lines in star-forming regions, for application to the NLR of AGN.
The code is based on {\sc HCm} \citep{hcm14}, which deals with optical emission lines for deriving O/H and N/O ratios in star-forming regions,
 and has been extended for use in the NLR of AGN in \cite{pm19b}.

 The above work has shown that when applied to star-forming galaxies, the abundances provided by the {\sc HCm} version using optical emission lines are in complete agreement with those provided by the direct method, while for AGNs they are in agreement with the expected metallicities at the centers of their host galaxies \citep{dors20}. 
This new version of the {\sc HCm-UV} code   is potentially useful for constraining the metallicity of the NLR of AGN up to very high redshift, 
and also provides a solution to the constraint imposed by using carbon emission lines to derive oxygen abundances, since it also estimates C/O. 

The paper is organised as follows: In Section 2, we describe the method for deriving chemical oxygen and carbon abundances in the NLR of AGN from their UV emission lines, based on the code {\sc HCm-UV}.
In the same section, we describe our grids of photoionisation models and
 we discuss their validity under different assumptions and for using different sets of emission lines.
In Section 3, we apply the method to a sample of compiled data on UV lines for NLR in AGNs, and in Section 4 we discuss the results and compare them with other published calibrations based on the same available emission lines.
Finally, in Section 5 we summarize our results and draw our conclusions.

\section{Description of the method}

The method discussed in this paper is the adaptation of the code {\sc HCm-UV}, originally developed for use with star-forming objects \citep{pma17}, to the NLR of AGN. 
The extension of different versions of the code for use on AGN has already been done for optical lines by \cite{pm19b} and for infrared lines by \cite{pd22}.
In this section, we discuss the model grids that the code uses to estimate O/H, C/O, and the ionisation parameter (log $U$) in these objects, as well as the defined observables based on the most typical UV emission line flux ratios that the code uses to derive these properties using a
Bayesian-like methodology. Finally, we discuss how different assumptions for the models and the emission lines used affect the results.

The version of the code described here (v. 5.0) along with other versions prepared for use in other spectral domains, is publicly available\footnote{In the webpage \url{http://www.iaa.csic.es/~epm/HII-CHI-mistry.html}}.

\subsection{Description of the models}

The entire grid of models used in HCm-UV was calculated using the photoionisation code {\sc Cloudy} v.17.01 \citep{cloudy}, which assumes a point-like central ionisation source surrounded by a gas distribution.
The models partially correspond to those of the optical version of the code described in \cite{pm19b},
taking into account a spectral energy distribution (SED)
with two components: one for the Big Blue bump peaking at 1 Ryd, and the other represented by a power law with spectral index $\alpha_x$ = -1 for the nonthermal X-rays. For the continuum between 2 keV and 2500 \AA \ we considered a power law with two possible values for the spectral index $\alpha_{ox}$=-0.8 and -1.2.
The first value fits better the emission line fluxes in tailored models by \cite{dors17}, while the second value fits better the average value found by \cite{miller11} in a sample of Seyfert-2 galaxies.
For the gas, we assumed a filling factor of 0.1 and a constant density of 500 cm$^{-3}$, as is typical in the NLRs around type 2 AGNs \citep{dors14}.
As discussed in \cite{pm19b} for the optical version of the code, assuming a higher value by 2$\cdot$ 10$^3$ cm$^{-3}$ does not lead to large deviations in the obtained results. Different grids were calculated considering the criterion of stopping the models and measuring the resulting spectrum as a function  of the fraction of free electrons ($f_e$) in the last zone (i.e., the outermost with respect to the ionising source). In some grids we considered a fraction of 98\% and in others a fraction of 2\%.
All chemical abundances were scaled to oxygen according to the solar proportions given by \cite{asplund09}, except for nitrogen and carbon, which are in solar proportion, but whose relation to oxygen was left in the models as an additional free input parameter.
Finally, we considered models with a dust-to-gas ratio that assumes the default value of the Milky Way, and also other models that do not consider the existence of dust mixed with the gas.
Although a larger variety in the assumptions of dust composition and proportions would be more realistic,
since these cannot be constrained from the input emission-line fluxes, these have not been more deeply explored in our grid of models,
avoiding an unnecessarily high number of models in our grids.

Overall, the models in each grid cover the range of 12+log(O/H) from 6.9 to 9.1 in bins of 0.1 dex, and values of log(C/O) from -1.4 to 0.6 in bins of 0.125 dex. In addition, all models consider values of log $U$ from -4.0 to -0.5 in bins of 0.25 dex.
This gives a total number of 5\,865 models for each of the resulting grids. 
Considering additionally the two possible values of $\alpha_{ OX }$ (-0.8 and -1.2), the two possible values of $f_e$
in the outermost zone (2\% or 98\%) and the existence or absence of dust grains mixed with the gas, we obtain a number of 8 different calculated grids.

\subsection{The {\sc HCm-UV} code adapted for AGNs}

The version of the code we describe here for its use for NLRs in AGNs follows a procedure similar to that 
 used in other versions for other spectral regions, such as the optical \citep{hcm14} or the IR \citep{jafo21}. In all cases, the code performs a Bayesian-like calculation that compares certain emission line fluxes and their errors with the results of the models in each grid.
Here we discuss the results obtained with the model grids described above. However, this new version of the code allows the user to use alternative grids as input models that have the same format as the default models.

The above comparison performed by HCm-UV to calculate the corresponding final mean values uses as weights for each model the $\chi^2$ values of some defined specific emission line ratios that are sensitive to the properties we want to derive, as described in \cite{hcm14}.
The uncertainties of the derived abundances and $U$ are calculated as the quadratic addition of the weighted standard deviation and the dispersion of the results after a MonteCarlo simulation, using the input errors as random perturbations around the nominal introduced values for each emission flux.

The list of UV emission lines allowed as input by {\sc HCm-UV}, for both SF and AGN, includes: 
Ly$\alpha$ at $\lambda$ 1216 \AA, \civ $\lambda$ 1549 \AA,
\oiii] $\lambda$ 1665 \AA, and \ciii] $\lambda$ 1909 \AA.
In addition, the code provides the ability to input optical emission lines H$\beta$ and [\oiii]$\lambda$ 5007 \AA \ to obtain estimates of abundances using emission line ratios that are sensitive to electron temperature, as is the case with 5007/1665 \citep{pma17}.

Unlike the previous version for star-forming galaxies, this version also includes the emission lines [\nv] at $\lambda$ 1239 \AA and \heii\ at $\lambda$ 1640 \AA, since the use of these lines in combination with the other carbon emission lines in the UV can be used to provide estimates of the total oxygen abundance in AGNs, as described in \cite{dors19}.
In any case, these two lines can also be used as input to the calculation of abundances in star-forming galaxies, since the presence of these lines can be considered a strong discriminating factor in the excitation of the gas when only massive stars are considered.

\begin{figure*}
\centering

\includegraphics[width=8cm,clip=]{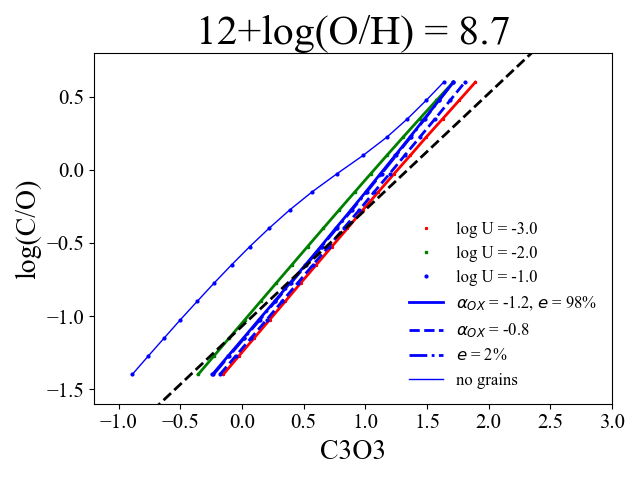}
\includegraphics[width=8cm,clip=]{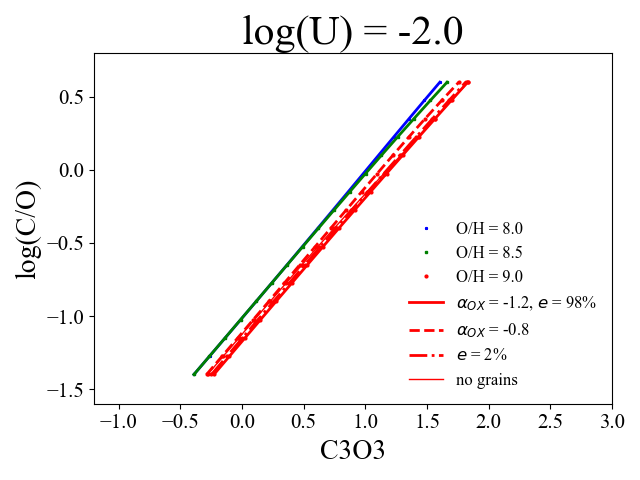}

\caption{Relationship between emission line ratio C3O3 and log(C/O) for model sequences with different values of input parameters (left for fixed 12+log(O/H) = 8.7, and right for fixed log $U$ = -2.0). The solid lines represent models for $\alpha_{ OX }$ = -1.2, $f_e$ in the outermost zone of 98\% and considering dust grains.
The other lines change only one parameter with respect to this sequence. The dashed line represents models with $\alpha_{ OX }$ = -0.8, the dot-dashed line represents models with $f_e$ of 2\%, and the thin line represents models without grains. 
 In the left panel, the black dashed line represents the linear fit for models of star-forming galaxies that are in \citet{pma17}.}

\label{c3o3-co}
\end{figure*}

\subsubsection{C/O derivation}
Following the same procedure as defined for SF objects, the code computes C/O in a first iteration, taking advantage of the fact that the emission line ratio C3O3 depends very little on $U$. This observable can be defined as already used by \cite{pma17} for star-forming galaxies to derive C/O:

\begin{equation}
\rm
C3O3 = \log \left( \frac{I(CIII] 1909)}{I(OIII] 1665)} \right)  
\end{equation}

In Figure \ref{c3o3-co} we show the relation between this parameter and C/O for model sequences with $\alpha_{ OX }$ = -1.2 and assuming $f_e$ = 98\% and grains at various log($U$) at a fixed 12+log(O/H) = 8.7 (left panel) and for various C/O values at a fixed log($U$) = -2.0 (right panel). We also show additional sequences of models with
$\alpha_{ OX }$ = -0.8, with $f_e$ = 2\%, and without dust grains to assess how varying these parameters affects this relation.

Figure \ref{c3o3-co} shows that despite a slight dependence on O/H, $U$ or $\alpha_{ OX }$, there is a well-defined linear relationship between the C3O3 parameter and C/O.
Only models without grains seem to predict a slightly lower slope compared to all other sequences. 
Compared to the linear relation derived in \cite{pma17} for models of star-forming galaxies, the sequences of the models for AGN are very close, albeit with a slightly lower slope, similar to that obtained for AGN without grains.

We conclude that C3O3 appears to be a robust indicator of C/O for AGN galaxies using UV emission lines, and that the code can subsequently be used for estimation.

\begin{figure*}
\centering

\includegraphics[width=8cm,clip=]{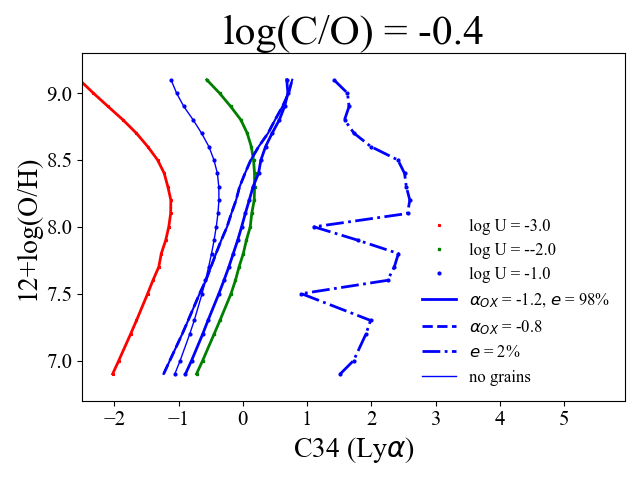}
\includegraphics[width=8cm,clip=]{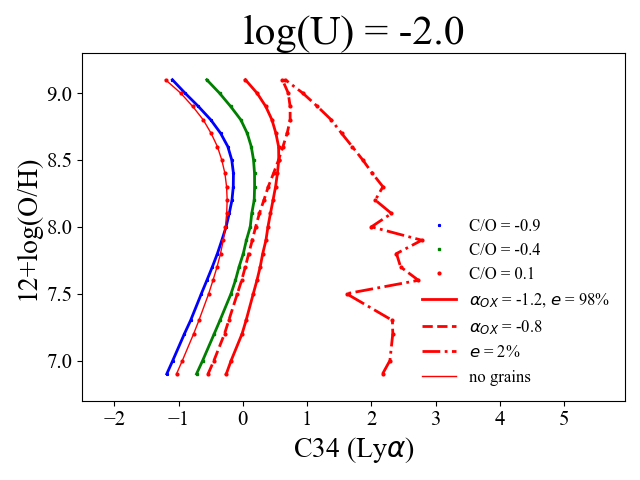}

\caption{Relationship between emission line ratio C34 using the Ly$\alpha$ line and total oxygen abundance for different model series at fixed log(C/O) = -0.4 (left panel) and log $U$ = -2.0 (right panel).
The solid lines represent models for $\alpha_{ OX }$ = -1.2, $f_e$ in the outermost zone of 98\% and considering dust grains.
The other lines change only one parameter in relation to this sequence.}

\label{c34-oh}
\end{figure*}

After estimating  C/O and its error, the code constrains the model grid to seek a solution for O/H and $U$ in a second iteration. This procedure ensures that C lines can be used without any prior arbitrary assumption about the relationship between O/H and C/O if the latter can be inferred.
If C/O cannot be estimated because some of the required lines are not available, the code assumes an expected relationship between O/H and C/O, as is the case for star-forming regions.
By default, the code assumes a solar C/N ratio and adopts the empirical relation between O/H and N/O derived in \cite{hcm14} for star-forming objects. However, in the current updated version of the code, other relationships can be considered.

\begin{figure*}
\centering

\includegraphics[width=8cm,clip=]{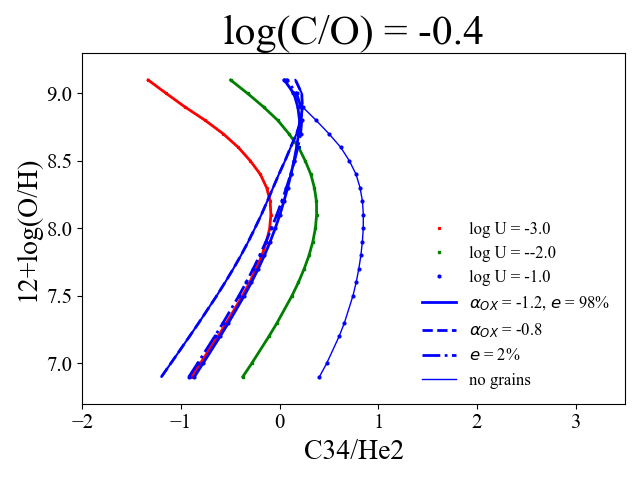}
\includegraphics[width=8cm,clip=]{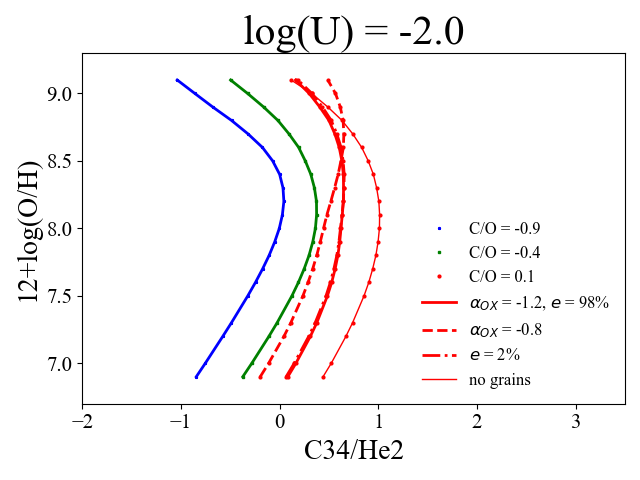}

\caption{Relationship between C34/He2 emission  line ratio and total oxygen abundance for different model sequences at fixed log(C/O) = -0.4 (left panel) and fixed log $U$ = -2.0 (right panel).
The solid lines represent models for $\alpha_{ OX }$ = -1.2, $f_e$ in the outermost zone of 98\% and considering dust grains.
The other lines change only one parameter with respect to this sequence.}

\label{c34he2-oh}
\end{figure*}

\begin{figure*}
\centering

\includegraphics[width=8cm,clip=]{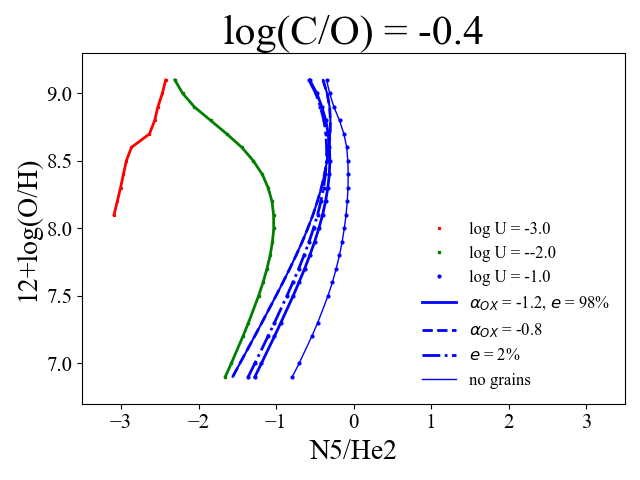}

\caption{Relationship between the emission line ratio N5/He2 and 12+log(O/H) for model sequences with different values of the input parameters at fixed log(C/O( = -0.4. The solid lines represent models for $\alpha_{ OX }$ = -1.2, $f_e$ in the outermost zone of 98\% and considering dust grains.
The other lines change only one parameter with respect to this sequence. The dashed line represents models with $\alpha_{ OX }$ = -0.8, the dot-dashed line represents models with $f_e$ of 2\%, and the thin line represents models without grains.
}

\label{N5He2-OH}
\end{figure*}

\subsubsection{O/H and $U$ derivation}

Among the various emission line ratios used in the second iteration to calculate the $\chi^2$ weights, the code uses parameter C34, defined as follows:
\begin{equation}
\rm
C34 = \log \left( \frac{I(CIV 1549) + I(CIII] 1909)}{I(H_{i})} \right)
\end{equation}
This parameter was also used for star-forming objects in \cite{pma17}.
In this case, I($Hi$) refers to the intensity of the hydrogen recombination emission lines. The closest and brightest object in this spectral range is Ly$\alpha$ at $\lambda$ 1216 \AA, but the code also allows the use of H$\beta$ at $\lambda$ 4861 \AA, although in this last case the extinction correction becomes much more critical.

In Figure \ref{c34-oh} we show the relation between this parameter, taking the Ly$\alpha$ intensity for the parameter, and O/H for some sequences of models from the grid at fixed log(C/O) = -0.4 (left panel) and log $U$ = -2.0.
 (right panel), with $\alpha_{ OX }$ = -1.2, $f_e$ = 98\% in the last zone and with dust grains. .

It is noted that the relationship between this parameter and the total oxygen abundance in certain sequences can be bivalued, since it increases for low values of O/H, remains almost constant and finally starts to decrease when the metallicity is very high.
This behaviour is in contrast to that observed in models for star-forming galaxies, where a monotonic increase in the parameter with metallicity is observed throughout the range studied, as discussed in \cite{pma17}.
However, the metallicity range where the turnover of the curve is observed depends on the assumed C/O and $U$, and can be nearly linear for log $U$ = -1.0. 
In fact, these two parameters have a very large influence on the relation of the parameter to the metallicity.

Despite the large dependence of this parameter on $U$ or C/O, the models predict a negligible deviation if we consider another harder SED with $\alpha_{ OX }$ = -0.8.
However, the assumption of absence of dust grains seems critical, since C34 shows a much more pronounced bivalued behaviour when no grains are considered, with very lower values at high metallicities, since Ly$\alpha$ is much more affected by dust extinction.

In addition, the inclusion of very low excitation zones also affects the behaviour of this parameter, since Ly$\alpha$ is much more strongly absorbed, so that C34 is much higher for the same O/H. 
This is consistent with the fact that Ly$\alpha$ 1216 $\lambda$ \AA \
is often absorbed by the neutral gas around the ionised gas nebulae and is therefore difficult to detect in many objects. 

\begin{figure*}
\centering

\includegraphics[width=8cm,clip=]{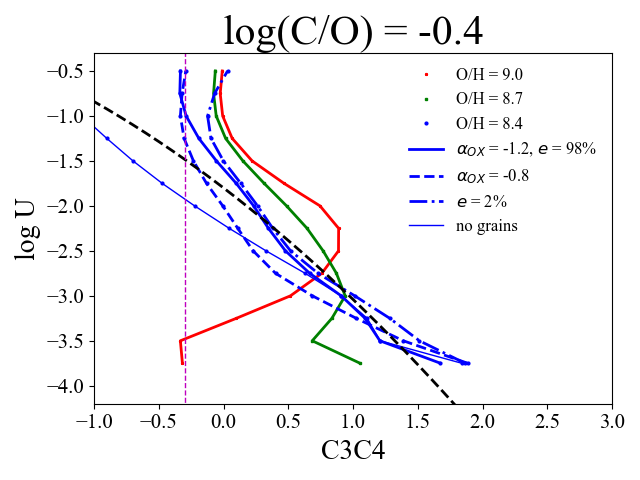}
\includegraphics[width=8cm,clip=]{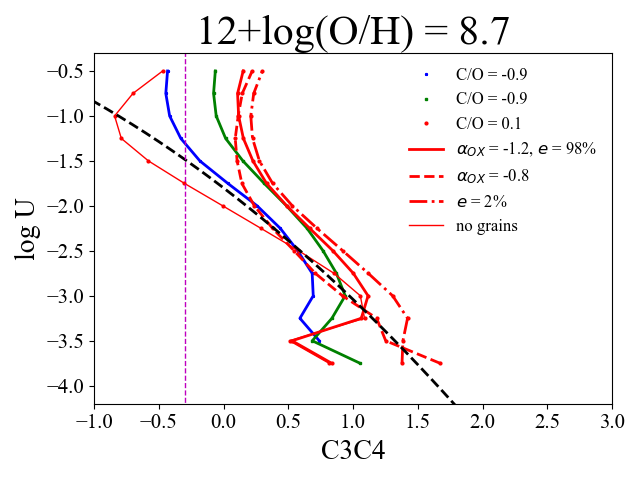}

\caption{Relationship between emission line ratio C3C4 and ionisation parameter for different model sequences at fixed log(C/O) = -0.4 (left panel) and fixed 12+log(O/H) = 8.7 (right panel). The solid lines represent models for $\alpha_{ OX }$ = -1.2, the ratio of free electrons in the outermost zone of 98\% and considering dust grains.
The other lines change only one parameter with respect to this sequence; the black dashed line represents the quadratic fit of Dors et al. (2019).
The vertical dashed magenta line marks the distance assumed for the C3C4 parameter to account for dust in the models for subsequent analysis.}

\label{c3c4-U}
\end{figure*}

An alternative is to define the same parameter as a function of the line \heii\ at $\lambda$ 1640 \AA, which is also used in \cite{dors19},
This takes advantage of the fact that this line is often well detected with high excitation in AGN and has less absorption by the surrounding neutral gas. 
The corresponding observable based on the same lines can be defined as follows,

\begin{equation}
\rm
C34He2  = \log \left( \frac{I(CIV] 1549) + I(CIII] 1909)}{I(HeII 1640)} \right)
\end{equation}
which is also used for the NLR of AGNs galaxies by \cite{dors14} named C43. However, in this work we refer to it as. 
C34He2 to distinguish it from our C34 parameter defined above.
The relation between C34He2 and O/H depending on the model is shown in Figure \ref{c34he2-oh} for the same model series at fixed C/O = -0.4 (left panel) and fixed log $U$ = -2.0 (right panel) and for a value of $\alpha_{ OX }$ = -1.2, $f_e$ = 98\% and with dust grains.
Furthermore, as in the previous plots, we show additional sequences with $\alpha_{ OX }$ = -0.8, with $f_e$ = 2\% in the last zone and also without dust grains.

As shown, the parameter has a similar trend to that observed for C34: it increases with increasing metallicity, although it enters saturation and begins to decrease at very high O/H values.
The metallicity at which this conversion is expected to occur is so dependent on $U$ that it is less pronounced for higher $U$ values.
This additional dependence of O/H on $U$ strengthens the basis of the procedure in the code by which these two parameters are calculated simultaneously in this second iteration.

The relation between C34He2 and O/H does not appear to change significantly when other, harder values of $\alpha_{ OX }$ are considered and, as expected, when a larger thickness of the nebula is assumed in the models, since \heii, unlike Ly$\alpha$, is not absorbed by neutral gas.
However, the models without grains predict much more different values of the parameter for the same O/H value, since the internal extinction of the gas affects the opacity of lines with high excitation, such as \civ] or \heii, leading to higher values of the parameter, in contrast to C34.

Another observable that the code can use in this second iteration is also defined and used by \cite{dors19} to derive O/H based on the highly excited emission lines that can be observed in the UV spectra of AGNs. It is the N5He2 parameter, which can be defined as follows,

\begin{equation}
\rm
N5He2  = \log \left( \frac{I([NV] 1239) }{I(HeII 1640)} \right)
\end{equation}

In Figure \ref{N5He2-OH} we show the relation between this parameter and O/H for model sequences at fixed C/O = -0.4 and varying log $U$, assuming a AGN SED with $\alpha_{ OX }$ = -1.2, a final zone with $f_e$ = 98\%, and grains. Additional sequences also consider other values for these input parameters.

Similar to C34 and C34He2, the N5He2 parameter also shows a strong dependence on $U$ and a bivalued relationship with oxygen abundance. The parameter has a lower value for lower $U$, but on the contrary, it does not appear to change the metallicity of the turnover point.

Moreover, since it depends on the lines with high excitation, it is more sensitive to the shape of SED, so it has lower values when we vary $\alpha_{ OX }$ = -0.8 and does not change significantly for a lower value of $f_e$ in the last zone of the models. Finally, the absence of dust grains is not negligible, but the difference is not as large as in the case of C34, due to the same cause.
Moreover, since N is also a secondary element, N5He2 has a strong dependence on N/O, which can be partially reduced by assuming that C/N remains unchanged in the models, although this is not fully justified since the stellar mass range of the nucleosynthetic production of these two elements is not exactly the same \citep{henry2000}, but can be similarly affected by processes of hydrodynamic gas exchange \citep{edmunds90}.
Changing N/O independently of C/O,  in the grids of models would allow us in principle to explore
in more detail the impact of N/O on the final metallicity derivation, but the absence of any UV emission-line ratio strictly dependent on N/O, does not allow us to incorporate it to the code and would unnecessarily enlarge the number of models in each grid. 

As commented, the dependence on C/O of the emission line ratios defined above to derive O/H  can be reduced by the prior determination of C/O in the first iteration using the C3O3 parameter. As for the excitation, the dependence of the various observables on it can also be reduced by the C3C4 emission line ratio defined as follows, 

\begin{equation}
\rm
C3C4  = \log \left( \frac{I(CIII] 1909)}{I(CIV]  1549)} \right)
\end{equation}

This ratio is already used by the code for star-forming objects and also by \cite{dors19} for the case of NLRs in AGNs.
This ratio was already proposed by \cite{davidson72} as an indicator of excitation in gas nebulae, although \cite{dors14} points out that it also depends on metallicity for low values of log $U$.

The relation of this emission line ratio with the ionisation parameter is shown in Figure \ref{c3c4-U} for different arbitrary model sequences at fixed log(C/O) = -0.4 (left panel) and fixed 12+log(O/H) = 8.7 (right panel) and assuming $\alpha_{ OX }$ = -1.2, $f_e$ = 98\% and with grains. As in the previous cases, models with harder SED are also shown with $\alpha_{ OX }$ = -0.8, with $f_e$ = 2\% and also without grains in the figures.

As can be seen, higher values of C3C4 indicate lower values of $U$ for most sequences, except for very high values of O/H, which show a bi-valued behaviour. This is a weaker effect compared to the relationship found for the O32 parameter for AGNs using the same models in the optical \citep{pm19b}, which exhibits bivalued behaviour at all values of O/H with a maximum value for log $U$ = -2.5. This monotonically decreasing relationship between C3C4 and $U$ makes it easier for the code to use the grid of models for all values of $U$.
This trend does not change noticeably when we consider a value for $\alpha_{ OX }$ = 0.8 or a lower $f_e$ for the termination criterion.

On the other hand, the C3C4 parameter is very sensitive to the presence of dust grains mixed with the gas. Neglecting the presence of dust in the models results in much lower values for C3C4. Indeed, the relationship that emerges from these models without dust is similar to the quadratic fit of \cite{dors19}, where the authors do not consider dust grains in their models.
For this reason, some authors, such as. \cite{nagao06} or \cite{dors19}, do not include dust in their models because many of the analysed objects have very low values of C3C4.
However, according to the model sequences shown in Figure \ref{c3c4-U}, models with dust cannot be excluded for values of the C3C4 parameter $ > $ -0.3, a threshold marked with a red line in the left panel of the figure.
In any case, other possible explanations for this very low values of C3C4 cannot be excluded and deserve future investigation, such as the possible existence of photon leaking, already
observed in some extreme star-forming galaxies (e.g. \citealt{schaerer2022}).

\begin{figure*}
\centering
\includegraphics[width=0.85\textwidth,clip=]{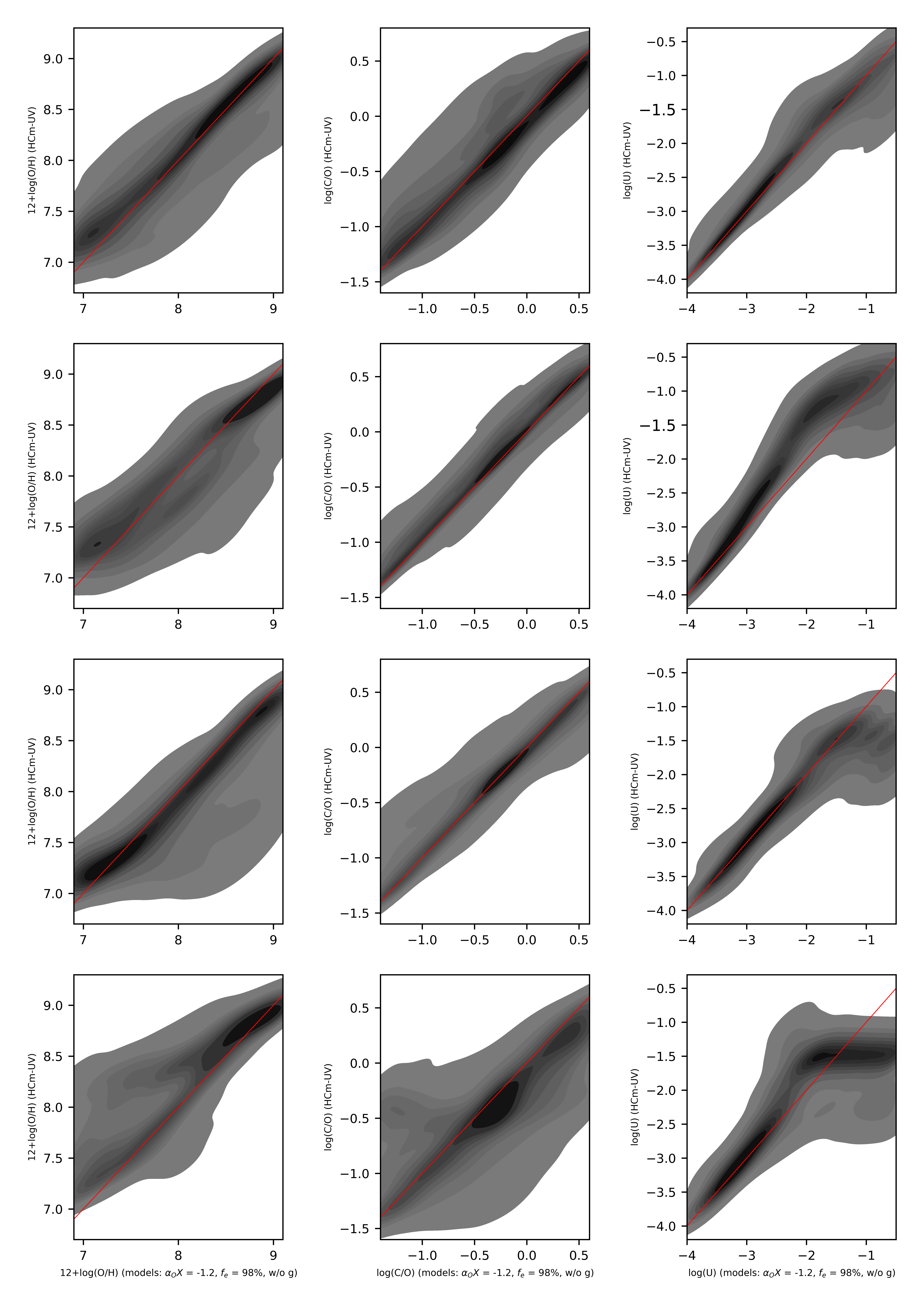}

\caption{Comparisons between the chemical abundances and $U$ assumed in the model grid and the same quantities calculated by {\sc HCm-UV} when the predicted emission lines are used as input. The panels in the left column show the comparison for 12+log(O/H), the middle column for log(C/O), and the right column for log $U$.
In all cases, {\sc HCm-UV} AGN models with $\alpha_{ OX }$ = -1.2, $f_e$ = 98\% and considering grains. In the panels of the first row, we show the results when the same conditions are assumed for the input; in the second, we change $\alpha_{ OX }$ to -1.2; in the third row, we consider $f_e$ = 2\%; and finally, in the bottom row, without dust grains. 
 The solid red line represents the 1:1 relation in all panels.}

\label{comp_mod}
\end{figure*}

\subsection{Testing the method with model emission-lines}

To verify the results of our model-based method, we used as input to the code the same emission line intensities predicted by the models to check that we obtain the same chemical abundances and ionisation parameters used for them. For this purpose, we also included a 10\% uncertainty in the predicted model-based fluxes to simulate the effect of an additional source of error on the estimates.
As explained above, the code uses the error of the flux as the standard deviation of a normal distribution of cases around the nominal value to perform a series of iterations that also reveal the associated uncertainty in the results.

In the panels of the first row in Figure \ref{comp_mod} we show the input abundances and $U$ assumed in the models of the grid for $\alpha_{ OX }$ = -1.2, a termination criterion when $f_e$ = 98\% is reached, and with dust grains. 
These values are compared with the results of the  {\sc HCm-UV} code when the same conditions are assumed and all lines accepted by the code are used. As can be seen, the agreement is excellent for most of the studied ranges (i.e., both the mean offset and the standard deviation of the residuals are less than 0.01 dex), so that both the abundances and the log($U$) values can be determined using only the emission line intensities. Only for 12+log(O/H) $ < $ 7.5 does the code tend to find solutions that are about 0.1 dex above the correct input value for each model, but still within the error limits (i.e., 0.17 dex).
For very high values of log($U$) (i.e. $ > $ -0.8), the code also tends to find results that are below the correct values, with deviations greater than the errors obtained (with a mean of 0.2 dex), but for the rest of the examined range there is agreement that is better than the associated errors.
 
We also investigate the potential impact in the results of changing the assumed conditions in the  library used by HCm-UV. Thus, we ran the code assuming the same conditions as above, but using as input  the emission lines predicted by the models when we assume a value $\alpha_{ OX }$ = -0.8, and leaving the other parameters equal to those assumed by the code. The corresponding comparisons are shown in the second line of Figure \ref{comp_mod}. 
In the case of O/H, the mean deviation is larger than the median uncertainty (0.18 dex) only at very low values, although there is also a systematic deviation for O/H $ > $ 8.0, in the sense that the code determines O/H values that are 0.1-0.2 dex below the correct values.
In contrast to O/H, the mean deviation for C/O is always lower than the obtained uncertainty (0.11 dex) in all ranges. This confirms that C/O can be determined more reliably than O/H when a different ionising SED is assumed. 
However, the $U$ estimate is more affected by the shape of the SED as it is significantly overestimated, especially in the range -3 $ < $ log$U$ $ < $ -1.

In the third line of Figure~\ref{comp_mod}, we show the comparisons when we change the truncation criterion in the models used as input to a fraction of free electrons = 2\% and leave the other parameters the same. Opposite to the input, the code still considers the same conditions as those assumed in the first case (i.e. with 98\% fraction of free electrons in the last model).
In this case, when the code assumes a different, higher $f_e$ value, the main offset occurs at high O/H values, where the code finds solutions about 0.3 dex below the correct input value, proving the importance of the relative intensity of the {\sc hi} lines to derive metallicity.
On the other hand, no large deviation is found for the C/O derivative, but in the case of $U$ the code systematically finds much lower values for log$U$ $ > $ -1.0.

Finally, in the bottom row of Figure \ref{comp_mod} we show the comparisons between the input values for models that do not consider dust and the values derived by the code assuming dust is present.
As shown, in this situation the code overestimates O/H with a mean offset larger than the mean uncertainty (i.e., 0.17 dex) for O/H $ < $ 8.0, but underestimates it for O/H $ > $ 8.5, albeit with a large spread in each case.
For C/O, the deviation between the solutions and the theoretical values is always within uncertainty (about 0.2 dex), except for very low values for C/O. It is also noted that there is a systematic tendency to find higher C/O values for high input values of C/O, but these deviations are always within errors.
Finally, in the case of $U$, we find better agreement, although the deviations are large for the highest values of $U$.

From these comparisons, it appears that the C/O ratio is more robust than the O/H ratio when the assumed conditions in the models vary relative to those assumed by the code. In the case of O/H, this is especially the case at high metallicities when the assumption about the presence of dust grains is incorrect, while for $U$ the largest discrepancies are found at very high values in all the comparisons tested. Finally, for log $U$ < -1.5 all discrepancies are found within the derived uncertainties.

\subsection{Results as a function of different sets of input emission-lines}

\begin{table*}
	\caption{Median deviations and root mean squares (RSME) of residuals between theoretical abundances and log($U$) values (AGN model inputs) and estimates from \textsc{HCm-UV} using different sets of emission lines and for different grids.}
	\label{offsets}     
	\centering          
	\begin{tabular}{llllllll}
\hline
\hline
		\textbf{Set of lines} & 		\textbf{Models} &\boldmath$\Delta_{OH} $ & \boldmath$\mathrm{RSME}_{OH}$ & \boldmath$\Delta_{CO} $ & \boldmath$\mathrm{RSME}_{CO}$ & \boldmath$\Delta_{U} $ & \boldmath$\mathrm{RSME}_{U}$\\ \hline 
\hline
		All lines & All & +0.01 & 0.21 & +0.02 & 0.14 & -0.01 & 0.26 \\ 
Ly$\alpha$, \nv], \civ], \heii, \oiii] $\lambda$ 1665 \AA, \ciii]  & All & +0.08 & 0.25 & -0.05 & 0.14 & +0.03 & 0.11\\
 \nv], \civ], \heii, \oiii] $\lambda$ 1665 \AA, \ciii]  & All & +0.08 & 0.25 & -0.05 & 0.14 & +0.03 & 0.11\\
 \nv], \civ], \heii, \ciii]  & All & +0.36 & 0.57 & -- & -- & +0.11& 0.34 \\ 
 \nv], \civ], \heii, \ciii]  & O/HC/O const. & +0.01 & 0.15 & -- & -- & +0.01& 0.13 \\
\civ], \heii, \ciii]  & All & +0.32 & 0.64 & -- & -- & +0.04 & 0.46 \\ 
\civ], \heii, \ciii]  & O/HC/O const. & +0.03 & 0.33 & -- & -- & -0.03  & 0.31 \\  
 \nv], \heii  & All & +0.27 & 0.66 & -- & -- & +0.08 & 0.48 \\ 
 \nv], \heii  & O/HC/O const. & +0.00 & 0.52 & -- & -- & -0.00 & 0.42 \\ 
\hline	\end{tabular}      
\end{table*}
Another important aspect in the evaluation of our code arises from the fact that it can provide solutions for different sets of emission lines.
To this end, in this subsection we compare the theoretical and resulting abundances, again using as inputs to the code different sets of emission line fluxes with an additional 10
The corresponding mean offsets and standard deviation of the residuals obtained from the code for O/H, C/O, and log $U$ compared to the theoretical values assumed by the models are shown in Table \ref{offsets},
for different combinations of emission lines that can be given as input to the code, and for different constraints on the model grid used.

As shown, when all lines allowed by the code are specified, the mean offsets of the results are better than the mean scatter in all cases. This comparison is the same as that in the first line of Figure \ref{comp_mod}.
This comparison is slightly worse when the optical line [\oiii] at $\lambda$ 5007 \AA, which is allowed by the code relative to
\oiii $\lambda$ 1665 \AA \ is not provided, but the results still agree better than the associated uncertainties.
A similar result is obtained if the emission line Ly$\alpha$ at $\lambda$ 1216\AA \ is removed, since \heii\ can be used instead. This has no influence on the C/O determination.

On the other hand, if the \oiii] $\lambda$ 1665 \AA \ is not provided, 
{\sc HCm-UV} cannot calculate the C3O3 ratio for the C/O estimate. In this case, very large deviations and errors result for both O/H and log $U$ when the full model grid is used for calculation. 
In this scenario, the code must assume a certain O/H-C/O ratio to get a better solution for O/H when C-lines are used. 
As shown in Table~\ref{offsets}, both the mean offsets and the standard deviation for O/H and log $U$ are much better under this additional assumption.
However, it is important to remember that this relation may be somehow arbitrary or valid for a particular galaxy population and not valid for a particular object.
In any case, even assuming such a relationship in the absence of the lines required for the previous derivation of C/O, much better results are obtained if the lines \ciii], \civ], and \nv] are used simultaneously with respect to \heii\, rather than calibrating them individually and in combination.

The above results again highlight the importance of a prior determination of the C/O content in order to obtain an accurate determination of the total metal content of the gas, and are in agreement with the results already obtained using the same approach for star-forming galaxies in \cite{pma17}.

\begin{landscape}
	\begin{table}
		\caption{List of UV fluxes for our sample of AGN. Column (1): name of the galaxy. Column (2): redshift. Column (3): spectral type. Columns (4)-(9): UV emission line fluxes and their errors in 1e$^{-14}$ erg/s/cm$^{2}$. Column (10): references: AS88 \citep{Allington_1988}, BER81 \citep{Bergeron_1981}, BES99 \citep{Best_1999}, BES00 \citep{Best_2000}, BOR07 \citep{Bornancini_2007}, BRE00 \citep{Breuck_2000}, BRE01 \citep{Breuck_2001}, CIM98 \citep{Cimatti_1998}, GK95 \citep{Gopal_1995}, KI93 \citep{Kinney_1993}, KRA94 \citep{Kraemer_1994}, LAC94 \citep{Lacy_1994}, LAC99 \citep{Lacy_1999}, LIL88 \citep{Lilly_1988}, LIL07 \citep{Lilly_2007}, LIN22 \citep{Lin_2022}, MAL83 \citep{Malkan_1983}, MAT09 \citep{matsuoka09}, MAT18 \citep{matsuoka18}, MCC90 \citep{McCarthy_1990}, MCC90b \citep{McCarthy_1990b}, MCC91 \citep{McCarthy_1991}, MCC91b \citep{McCarthy_1991b}, NA06 \citep{nagao06}, ONO21 \citep{Onoue_2021}, RAW96 \citep{Rawlings_1996},  ROB88 \citep{Robertis_1988}, ROT97 \citep{Roettgering_1997}, SIM99 \citep{Simpson_1999}, SNI86 \citep{Snijders_1986}, SPI95 \citep{Spinrad_1995}, STE99 \citep{Stern_1999}, STE02 \citep{Stern_2002}, SZO \citep{Szokoly_2004}, TAN22 \citep{Tang_2022}, THU84 \citep{Thuan_1984}, VM99 \citep{Villar_1999}, VM20 \citep{Villar_2020}, WAD99 \citep{Waddington_1999}. * stands for emission lines provided by their upper limit.$^{1}$Emission lines given as luminosity in units 10$^{43}$ erg$\cdot$s$^{-1}$. The complete version of this table is available at the CDS.}\label{Tabla_con_informacion}
		\centering
		\begin{tabular}{llllllllll}
			\textbf{Name} & \textbf{$z$} & \textbf{Type} & \boldmath$\mathrm{Ly\alpha}$ & \textbf{N\sc{v}1239$\AA$}  & \textbf{C\sc{iv}1549$\AA$}  & \textbf{He\sc{ii}1640$\AA$} & \textbf{O{\sc iii}]1665$\AA$} &  \textbf{C\sc{iii}]1908$\AA$} & \textbf{Ref.}\\  
		\textbf{(1)} & \textbf{(2)} & \textbf{(3)} & \textbf{(4)} & \textbf{(5)} & \textbf{(6)} & \textbf{(7)} & \textbf{(8)} & \textbf{(9)} & \textbf{(10)}\\ \hline 
		NGC 1068 & 0.004 & S2 & 713$\pm$100 & 224$\pm$41 & 520$\pm$80 & 187$\pm$29 & 22$\pm$8 & 240$\pm$35 & NA06,SNI86 \\
		I ZW 92 & 0.029 & S2 & 121.2$\pm$40.3 & - & 25.2$\pm$8.4 & 3.84$\pm$1.28 & - & - & KRA94 \\
		CDFS-901 & 2.578 & QSO & 3.71e$^{-3}\pm$6e$^{-5}$ & 6.5e$^{-4}\pm$8e$^{-5}$ & 1.97e$^{-3}\pm$1e$^{-4}$ & 1.87e$^{-4}\pm$9.3e$^{-5}$* & - & 3.3e$^{-4}\pm$9e$^{-5}$ & NA06,SKO04 \\
		CXO52 & 3.288 & QSO & 1.89e$^{-2}\pm$4e$^{-4}$ & 6e$^{-4}\pm$1.2e$^{-4}$ & 3.5e$^{-3}\pm$2e$^{-4}$ & 1.7e$^{-3}\pm$2e$^{-4}$ & 9e$^{-5}\pm$3e$^{-5}$ & 2.1e$^{-3}\pm$2e$^{-4}$ & NA06,STE02 \\
		S82-20$^{1}$ & 3.062 & QSO & 36.10$\pm$0.21 & 1.34$\pm$0.12 & 8.67$\pm$0.05 & 1.52$\pm$0.04 & 0.38$\pm$0.04 & 2.04$\pm$0.04 & LIN22 \\
		TXS J2334+1545 & 2.480 & HZRG & 3.1e$^{-2}\pm$5e$^{-3}$ & 7e$^{-4}\pm$3e$^{-4}$* & 1e$^{-3}\pm$1e$^{-3}$ & - & - & 8e$^{-3}\pm$3e$^{-3}$ & BRE00,BRE01 \\
		TN J1941-1951 & 2.667 & HZRG & 3.7e$^{-2}\pm$4$^{-3}$ & - & 9e$^{-3}\pm$1e$^{-3}$ & 4e$^{-3}\pm$1e$^{-3}$ & - & 2e$^{-3}\pm$1e$^{-3}$ & BOR07 \\
		COSMOS 05162 & 3.524 & QRG & 7.68e$^{-2}\pm$1.3e$^{-3}$ & 3.89$^{-3}\pm$1.95e$^{-3}$* & 1.19e$^{-2}\pm$8e$^{-4}$ & 2.6e$^{-3}\pm$3e$^{-4}$ & - & 4.05e$^{-2}\pm$6.5e$^{-3}$ & MAT18 \\
		COSMOS 10690 & 3.100 & QRG & 8.1e$^{-3}\pm$4.1e$^{-3}$* &4.5e$^{-3}\pm$2.3e$^{-3}$* & 2.25e$^{-3}\pm$6.8e$^{-4}$ & 8.2e$^{-4}\pm$4.1e$^{-4}$* & - & 1.7e$^{-3}\pm$5e$^{-4}$ & MAT18 	\\		
	\end{tabular}
	
\end{table}
\begin{table}
	\caption{Chemical abundances estimated from HCm-UV, using the grid of AGN models for $\alpha_{ OX }$ = -1.2 and the stopping criteria of 98\% of free electrons. Dust-free or grain-included models are selected based on their C3C4 value. Column (1): name of the galaxy. Column (2): oxygen abundance  (O/H) and its uncertainty. Column (3): carbon-to-oxygen (C/O) abundance ratio and its uncertainty. Column (4): ionisation parameter (U) and its uncertainty. Column (4): whether grids of models accounting for grains are employed. The complete version of this table is available at the CDS.}\label{Tabla_con_resultados}
	\centering
	\begin{tabular}{lllll}
		\textbf{Name} &  \boldmath$\mathrm{12+\log \left( O / H \right) }$ & \boldmath$\mathrm{\log \left( C / O \right) }$ & \boldmath$\mathrm{\log \left( U \right) }$ & \textbf{Grains} \\  
		\textbf{(1)} & \textbf{(2)} & \textbf{(3)} & \textbf{(4)} & \textbf{(5)} \\ \hline 
		NGC 1068 & 8.30$\pm$0.31 & -0.08$\pm$0.34 & -1.04$\pm$0.16 & No \\
		I ZW 92 & - & - & - & - \\
		CDFS-901 & 8.78$\pm$0.15 & - & -1.12$\pm$0.08 & No \\
		CXO52 & 7.79$\pm$0.26 & 0.24$\pm$0.26 & -1.39$\pm$0.21 & Yes \\
		TXS J2334+1545 & 8.82$\pm$0.10 & - & -2.61$\pm$0.16 & Yes \\
		TN J1941-1951 & 8.56$\pm$0.14 & - & -1.29$\pm$0.27 & No \\
		COSMOS 05162 & 8.77$\pm$0.14 & - & -1.16$\pm$0.16 & No  \\
		COSMOS 10690 & 8.68$\pm$0.09 & - & -0.84$\pm$0.11 & Yes 	\\		
	\end{tabular}
	
\end{table}
\end{landscape}

\section{Application to  galaxy samples}

We compiled from the literature strong UV narrow emission line fluxes for a sample of AGN consisting of Seyferts 2 (10), quasars (33), quiet radio galaxies (2), high-$z$ AGN (1), and high-$z$ radio galaxies (96). We take advantage of previous compilation work, (e.g. \citealt{nagao06, dors19}) of strong UV emission lines in AGN, namely \textsc{Nv}$\lambda $1239\AA, \textsc{Civ}$\lambda $1549\AA, \textsc{Heii}$\lambda$1640\AA \ and \textsc{Ciii}]$\lambda$1909\AA, but we also include information on the Ly$\alpha $ and [\textsc{OIII}]$\lambda$1665\AA \ emission lines when their measurements are given in the original references. Table \ref{Tabla_con_informacion} shows all the information we found for our sample, as well as the original references.

The Seyferts 2 (S2) sample consists of 9 objects taken directly from \citet{nagao06} and one additional galaxy, IZw\,92, from \citet{Kramer_1994}. The sample of high-$z$ radio galaxies (HZRG) was taken directly\footnote{From the original list of 167 sources, we omit 9 sources observed by \citet{matsuoka09} that provide better accuracy, and 79 sources that do not have enough UV spectroscopic information to be used as input to \textsc{HCm-UV}.} from the compilation presented by \citet{Breuck_2000}, supplemented by 8 sources observed by \citet{Bornancini_2007} and 9 galaxies analyzed by \citet{matsuoka09}. Our sample of quasars (QSO) consists of three different types of galaxies: Type II type quasars (11), 10 of them listed by \citet{nagao06} plus J142331.71$-$001809.1, recently analyzed by \citet{Onoue_2021}; extremely red quasars (21) from the compilation of \citet{Villar_2020}; and one intermediate Type I-II quasar, observed by \citet{Lin_2022}. We added to the above list of objects 2 quiet radio galaxies (QRG) observed by \citet{matsuoka18}, and a high-$z$ AGN (zAGN) whose broad and narrow components were identified by \citet{Tang_2022}.

Some of the compiled observations were already corrected for Galactic extinction in the original publications. These corrections were not very large (i.e. E(B-V) $\sim$ 0.01-0.1) due to their relative position on the sky, and do not   lead to  changes in the resulting emission-line fluxes larger than   the error limits.
On the other hand, we assumed that these objects were not corrected for internal extinction, so we performed this correction only for those objects with
 C3C4 $ > $ -0.3, which cannot  be considered as dust-free.
This criterion, shown in Figure \ref{c3c4-U}, is consistent with the predictions from models in which the absence of internal extinction is the only way to achieve the very low C3C4 values observed for certain objects \citep{nagao06,dors19}.
For higher C3C4 values, on the other hand, the absence of extinction does not seem to be justified in either the emission line flux correction  or the photoionisation models. Therefore, for the rest of the compiled galaxies, we assumed an average visual extinction of $A_V$ = 1 mag and an extinction law of \cite{ccm89}.
Anyway, we checked that the assumption of other common extinction laws does not imply a variation in the resulting chemical abundances
larger than the reported errors when only UV emission-lines are used as an input, as in our sample.
For the case of a simultaneous use of both optical and UV lines this correction would be more critical, but a more accurate estimate of the extinction correction could also be performed using the Balmer decrement.

Some of the listed fluxes for faint emission lines (e.g., \textsc{Nv}$\lambda $1239\AA, [\textsc{OIII}]$\lambda$1665\AA) are given as upper limits (noted in Table \ref{Tabla_con_informacion}). However, they are treated as real lines,
assuming the upper limit equivalent to 3$\sigma$, where the code considers as input the flux with a nominal value of 2$\sigma$ with an error of 1$\sigma$ (i.e. an upper limit of 3 is introduced as 2 $\pm$ 1), to reduce the probability that they are just noise,
although with a restriction for values $>$ 3$\sigma$ (i.e. larger than the upper limit) in our Monte Carlo simulations).

We applied the {\sc HCm-UV} code to this sample, assuming a AGN SED with $\alpha_{ OX }$ = -1.2 and a stopping criterion in the models for $f_e$ = 98\%.
The choice of these input parameters is arbitrary and corresponds mainly to the same conditions used for
the optical version of the code in \cite{pm19b}, with an $\alpha_{OX}$ value closer to the median derived by \cite{miller11}. .
Anyway, the effects of changing these parameters on the results are discussed in Section 2.
The only difference is that we only considered dust in the models for the objects whose lines were corrected for internal extinction, in agreement with the criterion based on C3C4, while for the rest, we used only the grids of the models without dust.
The resulting O/H, C/O, and log $U$ values are provided in electronic form along with the corresponding derived errors, as shown in Table \ref{Tabla_con_resultados}.
 
\begin{figure}
\centering

\includegraphics[width=8cm,clip=]{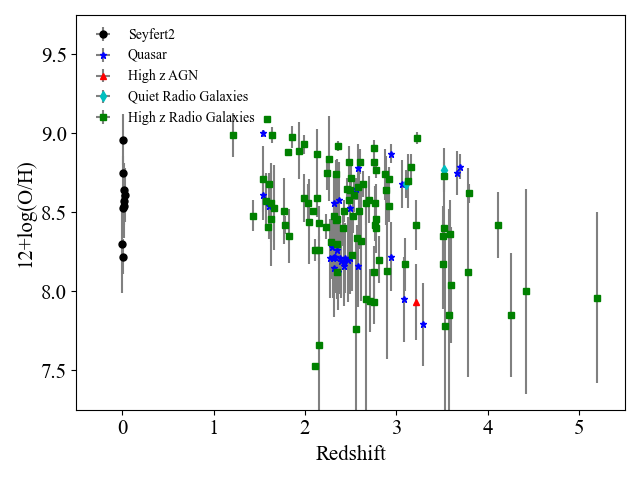}

\caption{Relationship between redshift and total oxygen abundances derived with the HCm-UV. Different symbols represent different object types in the sample: black circles for Seyfert 2, blue stars for quasars, red triangles for high-$z$ AGN, cyan diamonds for quiet-radiogalaxies, and green squares for high-$z$ radio galaxies.}

\label{OH-z}
\end{figure}

\section{Discussion}

\begin{table*}
	\caption{Mean resulting values obtained from {\sc HCm-UV} for O/H, C/O, and log $U$ in our sample of objects as a function of the galaxy type. We also list the mean redshift and the number of galaxies for which the calculation could be done to derive O/H and C/O, and the number of objects without dust grains.}
	\label{classes}     
	\centering          
	\begin{tabular}{llllllll}
\hline
\hline
		\textbf{Galaxy type} & 	\textbf{mean $z$}   & 	\boldmath$N_{OH}$ &  \boldmath$N_{wg}$ &   \textbf{Mean O/H}   &  \textbf{Mean log $U$}  &  \boldmath$N_{CO}$ &   \textbf{Mean C/O}   \\ \hline 
\hline
All  &  2.41   &   139   & 56 &  8.52   &   -1.48    &   26   & -0.50  \\   
Seyfert 2 & 0.015  &  9  & 6    &    8.56&   -1.40 &  1   &   0.06  \\
Quasar  &  2.51   &   32  &  20   &   8.35  &   -1.16  &  23   &  -0.53  \\
high-$z$ radio galaxies &  2.57  &  95  &   24   &      8.56  &  -1.65  &   1   &  -0.51  \\
High-$z$ AGN   &  3.21  &  1   &  1     &   8.03   &  -1.44  &   1   &   -0.27 \\
Quiet radio galaxies   &  3.31  &  2   &  1     &   8.84   &  -1.04  &   0   &   -- \\
\hline	\end{tabular}      
\end{table*}

\begin{figure}
\centering

\includegraphics[width=8cm,clip=]{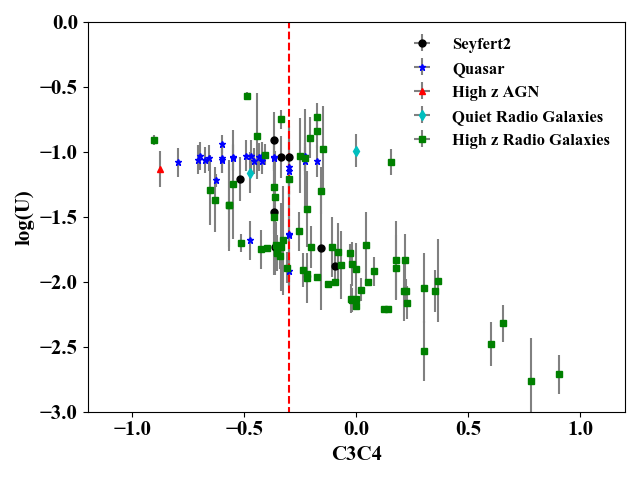}

\caption{Relation  between the C3C4 parameter and the ionisation parameter, log ($U$),  as derived using {\sc HCm-UV} for our sample of objects. The different symbols are the same as in Fig. \ref{OH-z}. The vertical red dashed line marks the limit below which objects are considered not having dust.}

\label{logU_sample}
\end{figure}

\subsection{O/H and U distributions}

In Figure~\ref{OH-z}, we show the derived metallicity in the compiled sample as a function of redshift for the objects marked for different categories.
We see no clear correlation with redshift, but a large scatter in the range of metallicity 7.77 $ < $ 12+log(O/H) $ < $ 8.97, with a mean of 8.52 ($\sim$ 2/3$\cdot Z_{\odot}$), which is slightly higher for the objects assumed to be dusty (i.e., 8.55) than for the objects we assume to be dust-free (8.46).
The main trend observed in this distribution would not be then very different assuming different input conditions in the models chosen to derive the abundances.

In Table \ref{classes} we show the mean O/H values obtained for each galaxy class. While these values do not allow us to draw firm conclusions, since these types are very unevenly populated and some of them could be very heterogeneous in their properties, they can provide some clues. 
For example, at very low $z$, Seyfert-2 galaxies have abundances around the mean of the whole sample, even if their mean is slightly subsolar. This is almost congruent with the sample of radio galaxies at high $z$, whose mean redshift is higher than that of the smaller Seyfert 2 subsample.
On the other hand, for the other types of galaxies at high $z$, very different values of O/H are found, as in the case of quasars, with a mean O/H value significantly lower (i.e., 12+log(O/H) = 8.35) than the mean of the whole sample. However, for the other three defined classes, the number of compiled objects is so small that no statistically significant conclusion can be drawn. 

In Figure \ref{logU_sample}, we show the relationship between the C3C4 parameter as obtained from the compiled emission line fluxes and the value of log $U$ as obtained from {\sc HCm-UV}. As described in previous sections, we considered all objects with C3C4 $ < $ -0.3 to be possibly dust-free, as discussed by \cite{nagao06} or \cite{dors19}. In table \ref{classes} we also give the fraction of objects in each category that fall within this range. As can be seen, more than half of the compiled objects in all categories are in this range and, could have in principle no dust, except for the radio galaxies with high $z$.

As for $U$, we find a relatively wide range of values from -2.7 $ < $ log $U$ $ < $ -0.6 with a mean $\log U=$-1.4, indicating that most objects have high excitation. Although we find a clear correlation between C3C4 and $\log U$, the objects assumed to not to have dust do not have a higher mean $U$ than the objects with dust. This is a consequence of the models predicting a lower C3C4 value in the absence of dust in order to obtain the same $U$ value.

In Table \ref{classes} we also report the mean log $U$ value in each category, which for Seyfert 2, as in the case of O/H, is very similar to the average values of the whole sample. In contrast, quasars show a much higher mean value, while radio galaxies with high $z$ show lower ionisation parameters on average.

\begin{figure*}
\centering

\includegraphics[width=11cm,clip=]{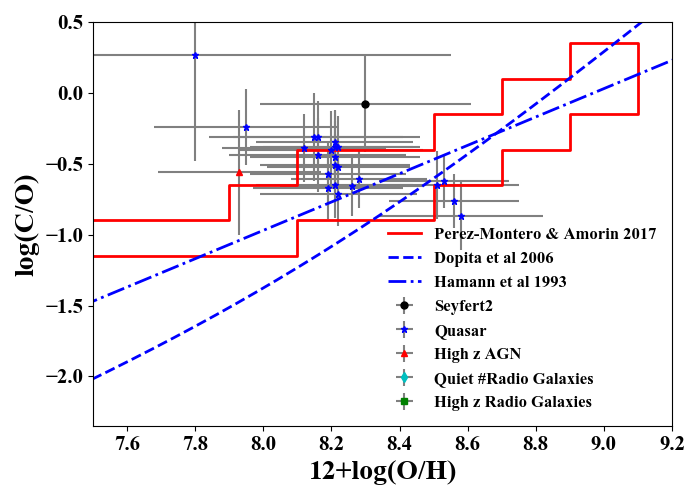}

\caption{Relationship between total oxygen abundance and carbon-oxygen abundance ratio derived using {\sc HCm-UV} for the assembled sample of NLR in AGN.
The red solid line encompasses the range covered by the models when the code in P\'erez-Montero \& Amor\'\i n (2017) assumes no prior derivation of C/O and a prior assumption about the ratio of O/H to C/O is required.
The blue dashed line corresponds to the relationship derived by Dopita et al. (2006) for \hii\ regions, and the dashed blue line represents the relationship derived by Hamann et al. (1993) for QSOs.}

\label{oh-co}
\end{figure*}

\subsection{C/O and its impact  on the O/H derivation}
For those objects in the sample for which the emission line ratio C3O3 was available, the code also estimated C/O.
Unfortunately, the number of objects in our sample for which this ratio is measurable is small (i.e., only 26 objects), most of which are quasars with an upper limit on the \oiii] $\lambda$ 1665 \AA \ line measurement, which could mean that the derived C/O values are only lower limits.
In Table \ref{classes} we also report the mean C/O values for each category, although they are not statistically significant for most classes because they could not be derived for many objects.

In Figure \ref{oh-co} we show the resulting O/H and C/O values with the corresponding errors for the objects for which these two ratios could be estimated simultaneously.
We note that most of the objects for which C/O could be derived have similar properties, since most of them are quasars all containing dust (according to the C3C4 criterion) and have relatively low O/H abundances in the range of 7.83 $ < $ 12+log(O/H) $ < $ 8.31.
The resulting C/O values for this subsample are in the range -0.84 $ < $ log(C/O) $ < $ 0.27, with a mean of -0.50 ($\sim$ 0.6$\cdot$ (C/O)$_{\odot}$), a somewhat lower proportion relative to the solar value than the corresponding O/H value for the entire sample, but significantly higher when compared to the mean for the subsample for which C/O could be measured (12+log(O/H) = 8.17 $\sim$ 0.3$\cdot Z_{\odot}$.

This relatively higher mean C/O value compared to the corresponding derived O/H value results in this subsample being on average above some of the commonly assumed relationships between O/H and C/O, as shown in Figure \ref{oh-co}. The sample is well above the relations given by \cite{hamann93} or \cite{dopita06}, and only a fraction of the objects are in the range considered by \cite{pma17}, remembering that the latter is only the conversion of the O/H-N/O relation assuming a solar C/N abundance ratio. 
It is not uncommon to find low-emission objects above these empirical or chemical model-based relations, since the relationship between metallicity and the ratio of abundance with a partial secondary origin relative to another with a primary origin has a large scatter (e.g., \citealt{pmc09}). The origin of such a scatter is related to several processes, including variations in star formation efficiency \citep{molla06} or gas exchange between galaxies and the surrounding IGM through hydrodynamic processes \citep{edmunds90,kh2005}, which likely affect the NLR in AGNs. 
This highlights the need for an alternative method to derive more accurate O/H abundances using UV carbon lines based on prior determination of C/O.

Indeed, the lack of prior determination of the C/O abundance ratio is one of the main sources of uncertainty we find in deriving chemical abundances from UV lines in both star-forming objects and AGN. In the absence of an emission line ratio sensitive to electron temperature, the calculation is based on the measured flux ratio of \ciii] and \civ\ lines relative to Ly$\alpha$ or \heii\,$\lambda$ 1640 \AA. 
The code {\sc HCm-UV} for star-forming objects presented in \cite{pma17} and the fit for AGNs presented in this work performs an initial iteration through the entire grid of assumed models to search for C/O using the observed C3O3 ratio, but we may wonder to what extent the derived final O/H values may vary if this previous step cannot be performed. 

We recalculated all O/H abundances in the subsample of 29 galaxies for which a prior derivation of C/O was possible, but this time without considering the \oiii] $\lambda$ 1665 \AA \ line. The latter means that the code cannot compute C/O and instead assumes the a priori expected relationship between O/H and C/O.
In this case, we obtain a mean O/H value of 8.66, which is 0.47 dex larger than the value obtained with a previous C/O derivation. Since most of the objects in our sample are quasars, the significantly lower O/H value derived for this category (see Table \ref{classes}) could have the origin explained above, and similar lower values cannot be discarded if an estimate of C/O can be given for the other object classes.

We also computed the O/H frequency assuming a restricted grid, taking the relationship between O/H and C/O proposed by \cite{dopita06} as a reference, since the new version of the code allows easy manipulation of this relationship, and we obtain an even larger mean frequency of 8.71. 
The 0.05 dex difference between the O/H frequencies derived from {\sc HCm-UV} when the program assumes the relationship given by \cite{pma17} or the relationship given by \cite{dopita06} or \cite{hamann93} in the absence of a prior C/O determination is also obtained for the entire object sample.

In summary, the very large variation between metallicities derived using C-lines from UV with or without prior C/O determination underscores the importance of measuring the C3O3 parameter to obtain accurate O/H abundances.

\begin{figure*}
\centering

\includegraphics[width=16cm,clip=]{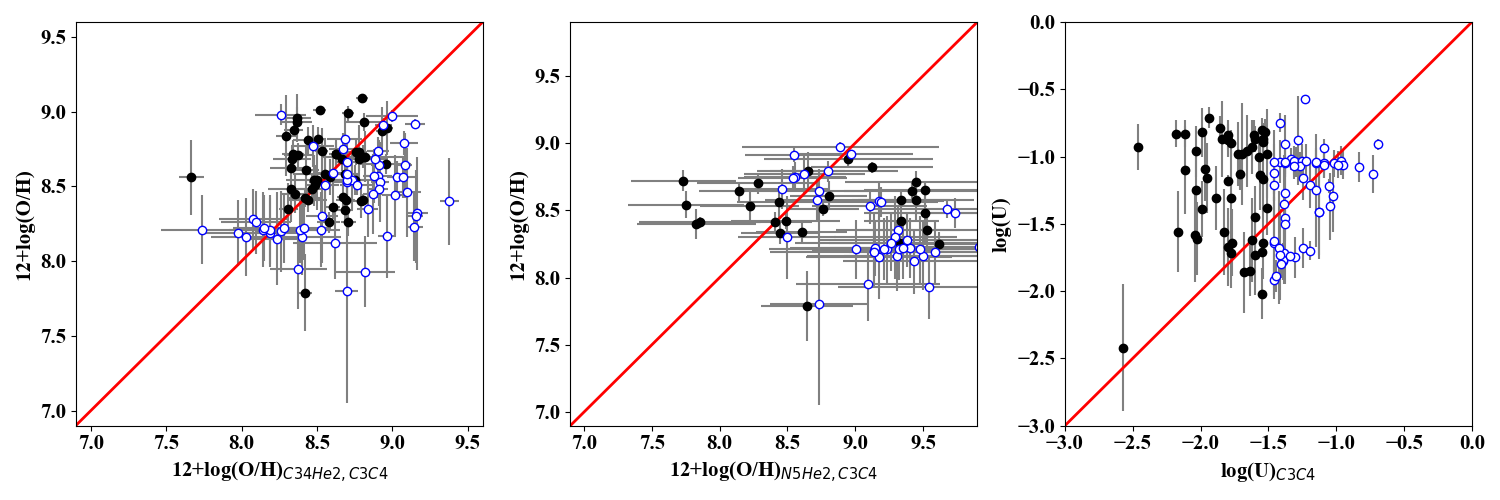}

\caption{Comparison between the resulting values from {\sc HCm-UV}, in the vertical axis, and those from different calibrations of Dors et al. (2019), in the respective horizontal axis, as applied to our galaxy sample. In all panels, filled symbols represent objects with assumed dust, while empty symbols represent objects without dust. In the left panel, the C34He2 and C3C4 parameter calibrations are used for O/H; in the middle panel, the N5He2 and C3C4 parameters are used for O/H; and in the right panel, the C3C4 parameter is used for log $U$.
The red solid lines represent the 1:1 relationship in all panels.}

\label{comp_d19}
\end{figure*}

\subsection{Comparison with other calibrations}

We use our sample to compare the total oxygen abundances and log$U$ derived from {\sc HCm-UV} with the results obtained with the model-based calibrations for AGNs with UV emission lines proposed by \cite{dors19}.

In the left panel of Fig. \ref{comp_d19}, we show the comparison for O/H obtained using the
 biparametric function proposed by \cite{dors19}, which is based on the combination of the C34He2 parameter with a correction of its dependence on $U$ using C3C4. We find agreement within the errors, both for objects with and without dust. Only for very large abundances (12+$\log$O/H$ > 9.0$) do we find a larger deviation due to the limit on {\sc HCm-UV} (i.e., the maximum O/H value is 9.1). 

Conversely, we find poor agreement when we compare the results of {\sc HCm-UV} with the abundances resulting from calibration based on N5He2 and C3C4 in \cite{dors19}, shown in the middle panel of Figure \ref{comp_d19}. As can be seen, the agreement is not good, and neither a clear correlation nor the range covered is similar for the two approaches. 
These strong observed differences cannot be attributed solely to the use of models with or without dust grains or to the assumption of different relationships between O/H and C/O, since these facts do not lead to such large differences, as discussed above. Instead, as we show in Table \ref{offsets}, they may be due to independent use of \ciii], \civ], and \nv] relative to \heii. 
Therefore, these results reaffirm the conclusion that the use of the N5HeII parameter alone is not advisable, but that, on the contrary, it must be used in combination with the other UV carbon lines.

In summary, the O/H derivation using these lines leads to very different results, even when the three lines are used and a C/O estimation was previously performed. 
For the sample of 26 galaxies with a prior determination of C/O using the C3O3 parameter, the mean difference between the O/H values obtained from {\sc HCm-UV} and the calibrations using the C34he2 and N5He2 parameters from \cite{dors19} is -0.2 dex and -0.9 dex, respectively. In contrast, better agreement is obtained when log $U$ is derived using the C3C4 emission line ratio, as shown in the right panel of Figure \ref{comp_d19}, which has an average offset of only 0.04 dex.

\subsection{Comparison with optical-based estimations}

Finally, we can compare the chemical abundances obtained from {\sc HCm-UV} using UV emission lines with those derived using optical emission lines from the version of the code described in \cite{pm19b} for a subset of galaxies with available information in both spectral regions. This is the case for NGC~1068, which was studied in \cite{nagao06}. Taking emission lines corrected for optical reddening relative to \hb, the code {\sc HCm} predicts 12+log(O/H) = 8.71 (8.59), assuming a value $\alpha_{ OX }$ = -0.8 (-1.2), which is very close to the value 12+log(O/H) = 8.68 (8.57) obtained by {\sc HCm-UV} using UV emission lines. 
For the other two galaxies in the sample, Mrk~3 and Mrk~573, also studied in \cite{nagao06}, optical spectra are available, but no UV \oiii] is reported at $\lambda$ 1665 \AA \, preventing a meaningful C/O estimate. Nevertheless, for these two galaxies, a solar N/O ratio was derived from the optical data, allowing us to assume a solar carbon-to-nitrogen ratio to compare the results from the optical and UV spectra. 
In the case of Mrk~3, we derived 12+log(O/H) = 8.58 (8.35) from {\sc HCm-UV} when we consider a $\alpha_{ OX }$ of -0.8 (-1.2), which is significantly lower than the values derived from the optical lines with {\sc HCm}, with 12+log(O/H) = 8.72 (8.59). 
For MRK~573, the differences are even larger, with 12+log(O/H) = 8.49 (8.26) from the UV and 8.79 (8.68) from the optical.

These differences are unlikely to be significant given i) the small number of objects for which simultaneous derivation of abundances in both the optical and UV can be performed, ii) the fact that the associated errors are of the same order of magnitude as these differences, iii) the strong dependence of the results on the extinction correction, especially in the UV, and iv) the assumption of a fixed C/N abundance ratio, which is not necessarily well justified. However, it is worth noting that the results are consistent overall and can be used as a reference for our sample of high-redshift galaxies.
	
\section{Summary and conclusions}

In this work we describe and implement the adaptation of the code {\sc HCm-UV} to the NLR of AGNs to derive oxygen abundances, the chemical carbon-oxygen abundance ratio, and ionisation parameters using UV emission line intensities.

According to our analysis based on photoionisation models covering different input features, the C3O3 emission line ratio turns out to be a robust indicator of C/O in AGNs, as it depends only to a very small extent on O/H, $U$, the shape of the considered SED, and the presence of dust grains mixed with the gas, which is of great importance in this spectral region.
The determination of C/O is not only important for the correct chemical interpretation of these objects, but also implies a much more accurate determination of metallicity based on the carbon UV emission lines.
This result is consistent with what has already been observed for SF objects in \cite{pma17}.

On the other hand, the determination of O/H and $U$, although much more reliable when a previous determination of C/O is provided using the C3O3 parameter, is
much more dependent on the assumed input conditions of the models, being particularly sensitive to the assumption of the presence or absence of dust grains mixed with gas. Only models without dust grains are able to reproduce the very low C3C4 values observed in many objects,
but the adoption of alternative matter-bounded geometry assumptions should also be explored. 

We applied our method to a wide range of data from the literature at different redshifts and we found a large scatter in the O/H and log $U$ distributions, but with a slightly subsolar mean and, as expected, high excitation. The derivation of C/O was only possible for a small subsample, mainly for quasars with values higher than expected for their derived metallicities, but with a large uncertainty because the emission line of \oiii] was inaccurately measured at $\lambda$ 1665 \AA. We verified that the prior determination of C/O for this subset dramatically affects the final derivation of their overall metallicity.

Finally, we compared our results with other methods based on the same compiled emission lines, such as the \cite{dors19} calibrations, and obtained very consistent results for the C34He2 parameter for O/H and C3C4 for log $U$, but not such good agreement for N5He2 for O/H. In this case, according to our models, the \nv]-emission line at $\lambda$ 1239 \AA \ should only be used in combination with the other lines, since it only provides an estimate of the very highly excited gas phase.

\section*{Acknowledgements}
This work has been partly funded by projects  Estallidos7 PID2019-107408GB-C44
(Spanish Ministerio de Ciencia e Innovacion),
and the Junta de Andaluc\'\i a for grant EXC/2011 FQM-7058.
This work has been also supported by the Spanish Science Ministry "Centro de Excelencia Severo Ochoa Program under grant SEV-2017-0709.
EPM also acknowledges  the assistance from his guide dog Rocko without whose daily help this work would have been much more difficult.
RA acknowledges financial support from ANID Fondecyt Regular 1202007.
R.G.B. acknowledges financial support from grant PID2019-109067-GB100. 




\bibliographystyle{mnras}

\bibliography{HCm-AGN} 








\bsp	
\label{lastpage}
\end{document}